\documentclass[3p,times]{elsarticle}

\usepackage{ecrc}

\volume{00}
\firstpage{1}
\journalname{Nuclear Physics A}

\runauth{Michael Benzke}

\jid{nupha}

\usepackage{amssymb}

\newcommand{\sh}[1]{#1\hspace{-5pt}/} 

\begin{document}

\begin{frontmatter}

%% use the tnoteref command within \title for footnotes;
%% use the tnotetext command for the associated footnote;
%% use the fnref command within \author or \address for footnotes;
%% use the fntext command for the associated footnote;
%% use the corref command within \author for corresponding author footnotes;
%% use the cortext command for the associated footnote;
%% use the ead command for the email address,
%% and the form \ead[url] for the home page:
%%
%% \title{Title\tnoteref{label1}}
%% \tnotetext[label1]{}
%% \author{Name\corref{cor1}\fnref{label2}}
%% \ead{email address}
%% \ead[url]{home page}
%% \fntext[label2]{}
%% \cortext[cor1]{}
%% \address{Address\fnref{label3}}
%% \fntext[label3]{}

%\dochead{}
%% Use \dochead if there is an article header, e.g. \dochead{Short communication}

\title{Gauge invariant definition of the jet quenching parameter}
\author{Michael Benzke}
\address{Physik-Department, Technische Universit\"at M\"unchen, James-Franck-Str. 1, 85748 Garching, Germany}
%\ead{michael.benzke@tum.de}

\begin{abstract}
We use the framework of Glauber extended Soft-Collinear Effective Theory to explicitly derive a gauge invariant expression of the jet quenching parameter $\hat{q}$. The effective theory approach offers a systematic power counting scheme at the Lagrangian level and allows for a consistent treatment of the relevant scales in the problem. Employing this approach in a covariant gauge scenario lead to an expression for $\hat{q}$ containing the expectation value of two light-cone Wilson lines. We find that in a general gauge, additional interaction terms in the Lagrangian have to be considered, leading to the introduction of transverse gauge links.
\end{abstract}


\end{frontmatter}

\section{Introduction}

Jet quenching refers to the phenomenon of a jet losing energy when transversing a medium, such as a quark-gluon plasma (QGP). Evidence for this new state of matter has been found in the heavy-ion collision experiments at RHIC and LHC. Jets, which are created in the early stages of the collision offer a suitable probe to analyze the properties of the QGP, since they are modified by the interaction with the medium. One of the observed effects due to the plasma is jet quenching, which has been measured at LHC and before at RHIC \cite{Chatrchyan:2011sx}.

Theoretically, the energy loss of a highly-energetic parton moving through a medium has been analyzed in several different formalisms \cite{Gyulassy:1993hr}. Recently, a new approach based on effective field theories (EFTs) was used to deal with one aspect of jet quenching, namely the transverse momentum broadening. This refers to a process, where the primary parton, which later fragments into a jet, interacts with the medium and thereby gains a certain momentum transverse to its original direction of motion.

The appropriate EFT for dealing with highly-energetic, almost light-like partons is Soft-Collinear Effective Theory (SCET) \cite{Bauer:2000ew}. In order to describe the interaction of the parton with the medium, the so-called Glauber mode was introduced into SCET in \cite{Idilbi:2008vm}. Based on this, momentum broadening was considered at leading order in the SCET power counting in \cite{D'Eramo:2010ak}. It was proven, that the jet quenching parameter $\hat{q}$, characterizing the effect of momentum broadening can be expressed in terms of the expectation value of two Wilson lines stretching along a light-like direction. 

However, the explicit proof only holds in a covariant gauge. Here, we will show how the picture changes in a general gauge. Specifically, we will explicitly derive an expression for $\hat{q}$ that holds in a general gauge by calculating the scattering of a highly-energetic parton in a background of Glauber gluons. It will turn out that, when considering a singular gauge, the Glauber supplemented SCET Lagrangian will contain additional leading power interaction terms, which give rise to a series of diagrams that lead to an expression valid in that gauge. This expression will contain transverse Wilson lines, connecting the previously known light-cone ones. Obtaining a gauge invariant expression is generally desirable, since it allows the computation using lattice gauge theory techniques and furthermore proves that the EFT calculation yields consistent results. The work presented here is based on \cite{Benzke:2012sz}.

\section{Theoretical Background}

\begin{figure}[ht] 
\center 
\includegraphics[width=67mm]{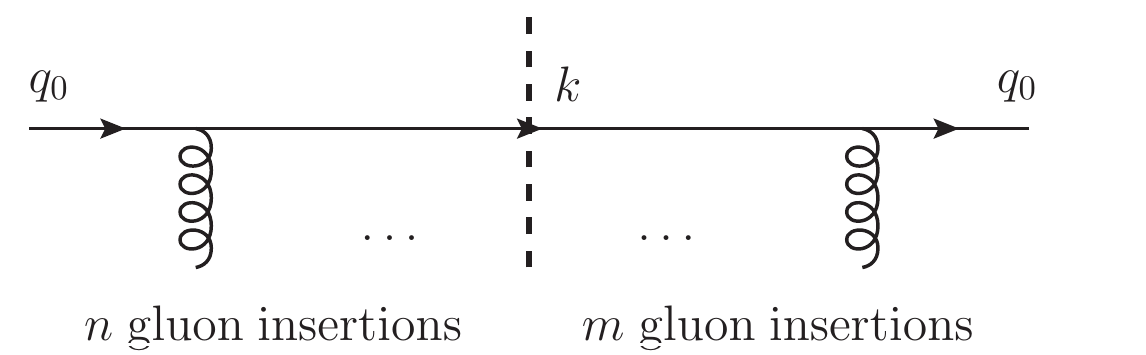} 
\caption{Cutting diagram contributing to the scattering amplitude. The dashed line is the cut.} 
\label{fig:amp} 
\end{figure}

The parameter $\hat{q}$ is defined as the average square transverse momentum with respect to the original direction of motion that the highly-energetic parton picks up while travelling a large distance $L$ through the medium. Following the notation of \cite{D'Eramo:2010ak} it can be written as
\begin{equation} 
\hat{q} = \frac{1}{L}\langle k_\perp^2\rangle = \frac{1}{L}\int\frac{d^2k_\perp}{(2\pi)^2}\,k_\perp^2 P(k_\perp)\,, 
\label{eq:qhat} 
\end{equation} 
where $P(k_\perp)$ is the probability for the hard parton to pick up a certain transverse momentum $k_\perp$. This probability can be related to the amplitude of a parton scattering on a number of constituents of the medium and thereby gaining a transverse momentum $k_\perp$. The squared amplitude is related via the optical theorem to the imaginary part of the forward scattering amplitude of a parton with initial and final momentum $q_0$. It is useful to first consider the forward scattering amplitude with $n$ interactions before and $m$ interaction after the cut, introduced by considering the imaginary part, and sum over every number of interactions at the end. The amplitude to calculate is depicted in fig. \ref{fig:amp}.
In order to obtain the probability $P(k_\perp)$ from the cut amplitude, we have to keep in mind that it is normalized with respect to the number of particles propagating through the medium and take into account the no-interaction part of the S-matrix \cite{D'Eramo:2010ak}.

Since we want to set up an EFT description of the process, we will now discuss the typical energy scales appearing in the problem. The large scale is set by the large momentum component $Q$ of the primary jet particle. A second, much smaller scale is the temperature $T$  or the general energy scale of the medium in non-thermal systems. This gives rise to a small dimensionless ratio $\lambda=T/Q\ll 1$. In order to characterize the relevant modes, we now introduce light cone coordinates, such that each four-vector is given by $q=(q^+,q^-,q_\perp)$, with $q^+=\bar n\cdot q$, $q^-= n\cdot q$, and the light-cone directions are given by $\bar n =\frac{1}{\sqrt{2}}(1,0,0,-1)$, $n = \frac{1}{\sqrt{2}}(1,0,0,1)$, such that $\bar n\cdot n = 1$. 

Here, we will consider a parton with initial momentum $q_0=(0,Q,0)$ travelling through the medium and picking up a transverse momentum of the typical medium scale, i.e., $Q\lambda$. We make no implicit assumptions concerning the production process of that parton, instead assuming it to be initially on-shell. After scattering on the medium constituents it will have a virtuality of order $k^2\sim Q^2\lambda^2$, which implies that its momentum scales like $Q(\lambda^2,1,\lambda)$. This mode will be called {\it collinear} and it will eventually fragment into the jet. Even though the fragmentation into other collinear particles is relevant, it significantly changes its energy and will not be considered in the calculation.

The typical representative of the medium whose momentum scales like $Q(\lambda,\lambda,\lambda)$ is called {\it soft}. In \cite{Hill:2002vw} it was suggested, that in the vacuum there are no interactions between collinear and soft particles at leading power in $\lambda$. However, the following derivation would in principle also work for soft gluon interactions.

The relevant contribution to jet quenching originates from interactions with gluons in the so-called {\it Glauber}-region with a momentum scaling of $Q(\lambda^2,\lambda,\lambda)$ or $Q(\lambda^2,\lambda^2,\lambda)$. They will not change the collinear nature of the jet parton, but significantly influence their transverse momentum component. SCET has been extended to include Glauber gluons in the context of a jet scattering on cold nuclear matter \cite{Idilbi:2008vm} and they have also been found necessary for the consistency of factorization proofs in certain processes like Drell-Yan \cite{Bauer:2010cc}.

With all the relevant modes at hand, we are now able to discuss the EFT appropriate for the calculation of the diagram in fig. \ref{fig:amp}. We restrict our analysis to the case of a highly-energetic parton that is a light quark. The SCET Lagrangian that deals with quarks with a large momentum component into the $\bar{n}$-direction and their interaction with gluons is given by 
\begin{equation} 
\mathcal{L}_{\bar{n}}=\bar{\xi}_{\bar{n}}\,i\sh{n}\,\bar{n}\cdot D\,\xi_{\bar{n}} 
+\bar{\xi}_{\bar{n}}\,i\sh{D}_\perp\,\frac{1}{2in\cdot D}\,i\sh{D}_\perp\,\sh{n}\,\xi_{\bar{n}}\,, 
\label{eqn:SCET} 
\end{equation} 
where $\xi_{\bar{n}}$ denotes the collinear quark field and the covariant derivative is $iD=i\partial+gA$. Since we are not considering collinear gluon emissions we will take the field $A$ to represent a Glauber gluon for now, which can be introduced as a background field. Furthermore, we can decompose the large component of the momentum of the collinear quark into the initial value $Q$ plus a small residual momentum due to the interaction with the plasma. Under these assumptions we can write the denominator in the second term of (\ref{eqn:SCET}) as $2in\cdot D\approx 2Q$, which makes the operator local, so we obtain the SCET Lagrangian relevant for the interaction with Glauber gluons
\begin{equation} 
\mathcal{L}_{\bar{n}}=\bar{\xi}_{\bar{n}}\,i\sh{n}\,\bar{n}\cdot D\,\xi_{\bar{n}} 
+\bar{\xi}_{\bar{n}}\,\frac{D_\perp^2}{2Q}\,\sh{n}\,\xi_{\bar{n}} 
+\bar{\xi}_{\bar{n}}\,i\frac{gG^{\mu\nu}_\perp}{4Q}\,\gamma_\mu\gamma_\nu\,\sh{n}\,\xi_{\bar{n}}\,,
\label{eqn:SCET2} 
\end{equation} 
where $D_\perp$ are two-component vectors such that $D_\perp^\mu D_{\perp \mu} = - D_\perp\cdot D_\perp$.
In order to determine which of the interaction terms contribute to the amplitude of fig. \ref{fig:amp}, we have to find the scaling of the components of the Glauber field itself.

Interestingly the scaling of the Glauber field depends on the gauge used. It turns out \cite{Idilbi:2008vm,Ovanesyan:2011xy} that $A_\perp$ is of the same or higher (depending on the source) order in the power counting as $A^+$ when using a covariant gauge. Therefore only the first term in (\ref{eqn:SCET2}) is of leading power in this case. In light-cone gauge ($A^+=0$) on the other hand $A_\perp$ is enhanced compared to $A_\perp$ in covariant gauge. This can be traced back to the factor $k_\perp^i/[k^+]$ appearing in the Fourier transform of the gluon propagator in light-cone gauge (the square brackets indicate an appropriate regularization for the singularity in the propagator). The factor $k_\perp^i/[k^+]$ is of order $1/\lambda$ for Glauber gluons. Hence, the second term of (\ref{eqn:SCET2}) is of leading power in light-cone gauge, while the first term trivially vanishes.

Another implication of the singularity in the momentum space propagator is that $A_\perp$ does not vanish for $x^-\to\pm\infty$ (when choosing a symmetric regularization). It is therefore convenient to decompose the gluon field \cite{GarciaEchevarria:2011md}
\begin{equation} 
A_\perp^i(x) = A^{\mathrm{cov},i}_\perp(x) + \theta(x^-)A_\perp^i(x^+,\infty,x_\perp)+\theta(-x^-)A_\perp^i(x^+,-\infty,x_\perp)\,,
\end{equation} 
where $A^{\mathrm{cov},i}_\perp$ is the non-singular part of the propagator and vanishes at $\pm\infty^-$. From the above power counting it follows, that the $A^{\mathrm{cov},i}_\perp$ part of the gluon field is suppressed compared to the field at $\pm\infty^-$ when using light-cone gauge.

Before starting the actual calculation we consider some properties of gauge fields at $\pm\infty^-$ \cite{Belitsky:2002sm}. The energy of a gauge field is proportional to the position space integral over the chromoelectric and chromomagnetic fields squared and is finite if both fields vanish at infinity. This requires that the field $A_\mu$ is a pure gauge at infinity, so we may write for an infinitesimal gauge transformation $1-ig\phi(x)$ that
\begin{equation} 
A_\mu(x)=\partial_\mu\phi(x)\,. 
\end{equation} 
In light-cone gauge $A^+=0$ this 
translates into a condition for the transverse components of the gauge fields at light-cone infinity, which can be solved by $\phi^\pm$ given by
\begin{equation} 
\phi^\pm(x^+,x_\perp) = - \int_{-\infty}^0 ds\, l_\perp\cdot A_\perp(x^+,\pm\infty,x_\perp+l_\perp s)\,, 
\end{equation} 
where we have chosen to integrate over a straight line going from $x_\perp-\infty l_\perp$ to $x_\perp$, $l_\perp$ being an arbitrary vector in the transverse plane. Finally, we note that the field strength tensor in the last operator in (\ref{eqn:SCET2}) also vanishes at $\pm\infty^-$.

\section{The Calculation of $\hat{q}$}

After the theoretical considerations of the last section, we are now in the position to calculate $\hat{q}$ in different gauges. Due to the scaling of the Glauber field the calculation in covariant gauge only needs to consider the interactions encoded in the first term of (\ref{eqn:SCET2}). This calculation was performed in \cite{D'Eramo:2010ak} and the result is given by
\begin{equation} 
P(k_\perp)=\int\,d^2x_\perp e^{ik_\perp\cdot x_\perp}\, 
\frac{1}{N_c} 
\left\langle \mathrm{Tr}\left\{W^\dagger[0,x_\perp]W[0,0]\right\}\right\rangle\,, 
\label{eqn:eramo} 
\end{equation} 
where $N_c$ is the number of colors, the brackets $\langle\dots\rangle$ denote the medium expectation value and the Wilson line $W$ is
\begin{equation} 
W[y^+,y_\perp]={\rm P}\,\exp\left[ig\int_{-\infty}^{\infty}\,dy^-A^+(y^+,y^-,y_\perp)\right]\,, 
\label{eqn:longWilson} 
\end{equation} 
with ${\rm P}$ the path ordering operator and in the limit $L\to\infty$. This was first derived in \cite{CasalderreySolana:2007zz} using a different approach.

In light-cone gauge, only the second term in (\ref{eqn:SCET2}) contributes to the amplitude of fig. \ref{fig:amp} at leading power. It gives rise to vertices connecting a collinear quark to the transverse component of either one or two gluons. A scattering amplitude with $n$ gluons attached can therefore consist of any combination of one- or two-gluon vertices, that add up to the correct amount of gluons. For the calculation of the amplitude we make use of an iterative approach \cite{Benzke:2012sz}. We find that the resummation of an infinite number of Glauber gluon interactions in light-cone gauge gives rise to transverse Wilson lines $T$ at light-cone infinity $\pm\infty^-$. They are given by \cite{Idilbi:2010im}
\begin{equation} 
T(x_+,\pm\infty,x_\perp) = {\rm P}\,\exp\left[-ig\int_{-\infty}^0 ds\; l_\perp\cdot A_\perp(x^+,\pm\infty,x_\perp+l_\perp s)\right]\,. 
\label{eqn:TWilson} 
\end{equation} 

\begin{figure}[ht] 
\center 
\includegraphics[width=90mm]{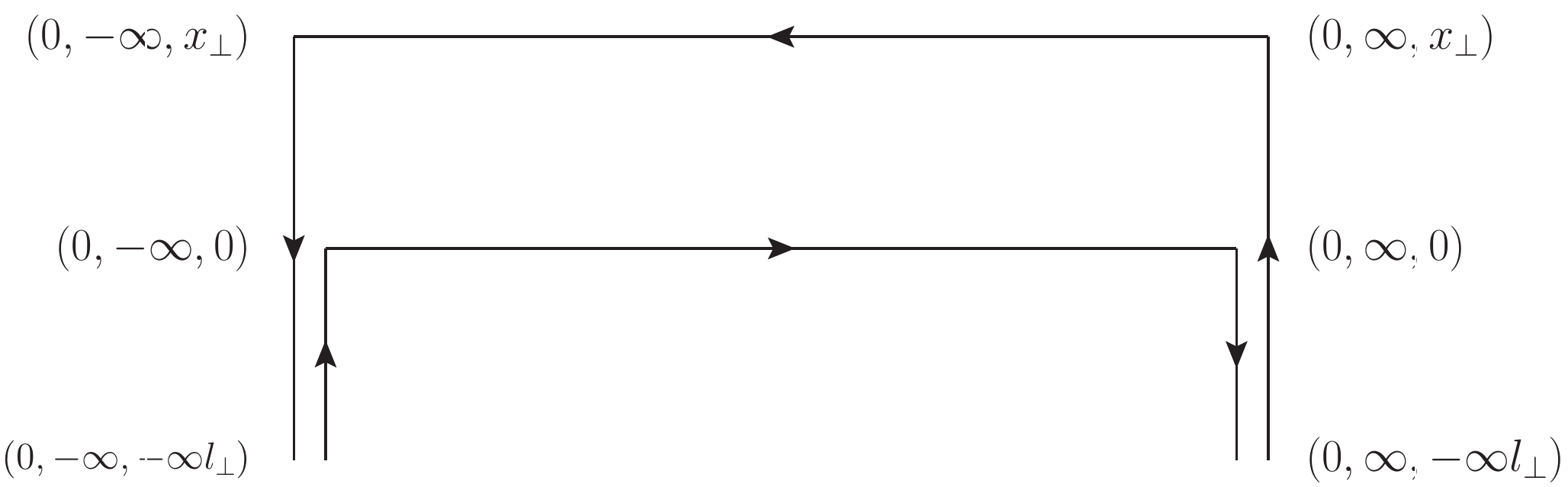} 
\caption{Wilson lines appearing in (\ref{eqn:P}).  
Horizontal lines stretch along the $\bar{n}$ direction, vertical lines along the vector $l_\perp$ in the transverse plane.}
\label{fig:p2} 
\end{figure}

Combining the contributions from all terms in (\ref{eqn:SCET2}) in the calculation of the scattering amplitude, we can write the gauge invariant result for the probability $P(k_\perp)$ in (\ref{eq:qhat}) as
\begin{equation}
P(k_\perp)=\int\,d^2x_\perp\, e^{ik_\perp\cdot x_\perp} 
\frac{1}{N_c}\left\langle\mathrm{Tr}\left\{ 
T^\dagger(0,-\infty,x_\perp)\, W^\dagger[0,x_\perp]\, T(0,\infty,x_\perp)\; 
T^\dagger(0,\infty,0)\, W[0,0]\, T(0,-\infty,0) \right\}\right\rangle \,,
\label{eqn:P}
\end{equation}
where the Wilson lines $W$ and $T$ are defined in (\ref{eqn:longWilson}) and (\ref{eqn:TWilson}), respectively. The path of the Wilson lines is depicted in fig. \ref{fig:p2}.
It is easy to show explicitly that this expression is gauge invariant \cite{Benzke:2012sz}. Some care has to be taken due to the fact that operators on the lower path of fig. \ref{fig:p2} are time ordered, while the ones on the upper path are anti-time ordered.
This concludes the explicit EFT derivation of $\hat{q}$ which agrees with similar expressions conjectured in \cite{D'Eramo:2010ak,Liang:2008vz}. The fully gauge-invariant expression not only allows for computations in any gauge, but also opens the way for the use of lattice data in the evaluation of $\hat{q}$, as suggested in \cite{CaronHuot:2008ni,Majumder:2012sh}.

\bibliographystyle{elsarticle-num}

\end{document}